\documentclass[prd,twocolumn,floatfix,superscriptaddress]{revtex4-1}
\usepackage{dcolumn,amsmath}
\usepackage{graphicx}
\usepackage{braket}
\usepackage{bm}
\usepackage{amssymb}
\usepackage{hyperref}
\usepackage{multirow}
\usepackage{footnote}
\usepackage{subcaption}
\usepackage{ragged2e}
\usepackage{xcolor}
\usepackage{pdfcomment}
\usepackage[justification=justified,labelfont=bf,singlelinecheck=off]{caption}
\usepackage{threeparttable,booktabs}  
\usepackage{subcaption,siunitx,booktabs}
\usepackage{caption,afterpage,tabularx}
\usepackage{diagbox}
\usepackage{mathrsfs}
\usepackage{physics}
\usepackage{cancel}
\usepackage{amsmath}
\usepackage{amssymb}

\newcommand{\JSU}{
School of Physics and Electronic Engineering,
\\Jiangsu University, Zhenjiang, 212013 Jiangsu, China\\
}

\newcommand{\MUST}{State Key Laboratory of Lunar and Planetary Sciences, 
\\Macau University of Science and Technology, 999078 Macao, China}

\begin{document}

\title{Dineutron decay into sterile anti-neutrinos in neutron stars and its observable consequences}
\author{Yongliang Hao}
\affiliation{\MUST}
\affiliation{\JSU}
\author{Dongdong Ni}
\email{ddni@must.edu.mo}
\affiliation{\MUST}


\date{\today}

\begin{abstract}
In some extensions of the Standard Model (SM), two neutrons are allowed to decay into two sterile anti-neutrinos ($nn \rightarrow \bar{\chi}\bar{\chi}$) via new scalar bosons. This process violates both the baryon number ($\mathcal{B}$) and the lepton number ($\mathcal{L}$) by two units but conserves their difference $(\mathcal{B}-\mathcal{L})$. Neutron stars contain a large number of neutrons and thus the $nn \rightarrow \bar{\chi}\bar{\chi}$ process can be greatly enhanced inside a neutron star. This process could result in non-trivial effects that are different from the SM predictions and can be explored through astrophysical and laboratory observations. Furthermore, a large number of sterile antineutrinos, which may be dark matter candidates, can be emitted from the interior of the neutron star. The properties of the emitted particles show a particular pattern that can be uniquely determined by the mass and radius of the neutron star. In addition, the dineutron decay may contribute to the orbital-period change of the binary systems containing neutron stars. We analyze the possibility to constrain the mass of the new scalar bosons using the observations of the binary's orbital-period changes. It is found that the mass of the new scalar bosons is roughly restricted in the range from 1 TeV to several TeV, which is possibly within the reach of direct searches at the LHC or future high-energy experiments. The joint analysis which combines the astrophysics and particle phenomenology could provide an excellent opportunity for the study of the new physical effects beyond the SM.

\end{abstract}

\maketitle

\section{Introduction}
The fundamental particles and three of the fundamental interactions, such as the strong, weak, and electromagnetic interactions, can be properly described by the Standard Model (SM) of particle physics \cite{zyla2020review}. The discovery of the SM-like Higgs boson \cite{aad2012observation,chatrchyan2012observation,chatrchyan2013observation} indicates one of the major achievements of the SM. However, there are still many open questions that cannot be well explained by the SM. One of the great challenges is the matter-antimatter asymmetry, which refers to the observed excess of matter over antimatter in our universe \cite{zyla2020review}. According to the Big Bang theory \cite{gamow1946expanding,alpher1948origin}, equal amounts of matter and antimatter should have been created in the early universe. It seems that, so far, the initial antimatter have disappeared and such a disappearance remains a puzzle \cite{zyla2020review}.

Baryon number ($\mathcal{B}$) and lepton number ($\mathcal{L}$) are usually considered as accidental symmetries in the framework of the SM \cite{cerdeno2020impact}. Some non-perturbative effects may violate the $\mathcal{B}$, $\mathcal{L}$, or ($\mathcal{B}+\mathcal{L}$) symmetries, but the difference ($\mathcal{B}-\mathcal{L}$) is still conserved \cite{hooft1976symmetry,hooft1976computation,arnold2013simplified,ellis2016search,kuzmin1985anomalous}. For example, some non-perturbative sphaleron processes may violate the $(\mathcal{B}+\mathcal{L})$ symmetry but preserve the $(\mathcal{B}-\mathcal{L})$ symmetry \cite{kuzmin1985anomalous}. $\mathcal{B}$-violation is one of the three conditions presented by Sakharov to explain the observed matter-antimatter asymmetry \cite{sakharov1967violation}. Furthermore, $\mathcal{B}$-violation plays a critical role in the construction of the extensions to the SM as it is implemented as an important feature in many new physics models \cite{zyla2020review}.

The dineutron decay into dilepton final states have been mathematically predicted by many new physics models \cite{heeck2020inclusive,girmohanta2020improved,girmohanta2020baryon,he2021eft} and intensively explored in a wide variety of experiments \cite{bernabei2000search,araki2006search,anderson2019search,collaboration2022improved}. From the experimental aspects, the limits on the lifetimes for the decay modes with invisible (or weakly interacting) final states have been reported by various experiments, such as LNGS ($1.2 \times 10^{25}$ yr \cite{bernabei2000search}), KamLAND ($1.4 \times 10^{30}$ yr \cite{araki2006search}), SNO+ ($1.3 \times 10^{28}$ yr \cite{anderson2019search}, $1.5 \times 10^{28}$ yr \cite{collaboration2022improved}), and etc. Among various processes, the dineutron decay into two sterile antineutrinos ($n n  \rightarrow \bar{\chi} \bar{\chi}$) has many interesting signatures, distinguishing it from other decay modes. Here, $\chi$ ($\bar{\chi}$) denotes the sterile neutrino (antineutrino), which may have a non-zero lepton number $\mathcal{L} \equiv 1$ ($-1$) and does not exist in the framework of the SM. The $n n  \rightarrow \bar{\chi} \bar{\chi}$ process violates the $\mathcal{B}$, $\mathcal{L}$, and ($\mathcal{B}+\mathcal{L}$) symmetries but conserves ($\mathcal{B}-\mathcal{L}$) symmetry. The sterile antineutrino barely interacts with the SM particles and could be dark matter candidates \cite{abazajian2001sterile}. From the theoretical aspect, some new physics models have been constructed with ($\mathcal{B}-\mathcal{L}$) symmetry. For instance, the model based on the group $SU(3)_c \times SU(2)_L \times SU(2)_R \times U(1)_{B-L}$ \cite{pati1974lepton,mohapatra1975left,mohapatra1975natural,senjanovic1975exact} implements ($\mathcal{B}-\mathcal{L}$) as a conserved quantity \cite{davidson1979b,mohapatra1980local}, rather than $\mathcal{B}$ alone or $\mathcal{L}$ alone. This model can be further accommodated in the models with additional symmetries for a unified description of quarks and leptons, such as the Pati-Salam model \cite{pati1974lepton,pati1975erratum} and its variants \cite{davidson1979b,mohapatra1980local} based on the $SU(4)_c \times SU(2)_L \times SU(2)_R$ group. Therefore, the $nn \rightarrow \bar{\chi}\bar{\chi}$ process can serve as a promising probes for such new physics models. However, the transition rate for this process is highly suppressed by the new-physics energy scale. Furthermore, this process is featured with the decay of two neutrons into two back-to-back energetic sterile antineutrinos. Since the sterile antineutrinos barely interact with the ordinary matter, they are almost invisible in the present detectors. These factors impose a great challenge for the detection of the sterile neutrinos in the laboratory experiments.

Neutron stars are one of the densest objects in our universe and can serve as a neutron-rich environment where many interesting processes and phenomena associated with neutrons occur \cite{berryman2022neutron}, making it possible to search for $\mathcal{B}$-violating effects through astrophysical observations. $\mathcal{B}$-violating effects can be induced by high-dimension operators and thus are highly suppressed by the new physics energy scale. Specifically, the $nn \rightarrow \bar{\chi}\bar{\chi}$ process can be mediated by the new scalar bosons through the interactions described by dimension-12 operators. The direct searches for new particles at the LHC shows that no significant evidence of such new scalar bosons beyond the SM has been found so far, suggesting that the new physics energy scale tends to be so large that direct laboratory detection might be inappropriate through the present experimental techniques. By contrast, neutron stars contain a large number of neutrons and the dineutron decay inside them can emit a large number of sterile antineutrinos. Due to this process, the neutron star would gradually lose mass and change its properties, bringing in observable effects in astrophysical observations.

In this work, we organize our discussions as follows. To begin with, we review the new physics models with additional new scalar bosons that lead to the $nn \rightarrow \bar{\chi}\bar{\chi}$ process. Next, based on such models, we estimate the decay rate for the $nn \rightarrow \bar{\chi}\bar{\chi}$ process. After that, we briefly review the structure of neutron stars and the equation of state for the neutron-star matter. Then, we transfer our attention to the observable consequences of the $nn \rightarrow \bar{\chi}\bar{\chi}$ process on the properties of neutron stars, such as particle emission and orbital-period change. In the following discussions, unless otherwise specified, we will adopt the natural units (i.e. $c \equiv 1$, $\hbar \equiv 1$).

\section{The model}

\begin{figure}[b] 
\centering
\includegraphics[scale=0.99, width=0.65\linewidth]{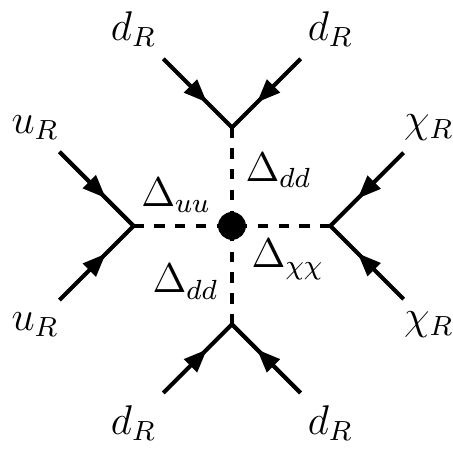}
\caption{A possible diagram for the dineutron decay into two sterile antineutrinos ($n n\rightarrow \bar{\chi} \bar{\chi}$) mediated by the new scalar bosons at the tree level.
}
\label{dineutron}
\end{figure}

Fig. \ref{dineutron} shows a possible diagram at the tree level for the dineutron decay into two sterile antineutrinos ($n n\rightarrow \bar{\chi} \bar{\chi}$) mediated by the new scalar bosons, namely diquarks and dileptons \cite{mohapatra1982hydrogen,mohapatra1983spontaneous}. This is not the only diagram that is responsible for the $n n\rightarrow \bar{\chi} \bar{\chi}$ process. Such a process can also be mediated by diquarks and leptoquarks. The new scalar bosons can be accommodated in some new physics models with additional symmetries, such as the left-right symmetric (LRSM) model based on the group $SU(3)_c \times SU(2)_L \times SU(2)_R \times U(1)_{B-L}$ \cite{pati1974lepton,mohapatra1975left,mohapatra1975natural,senjanovic1975exact}. The LRSM can be further embedded in some (partially) grand unified models with higher symmetries, such as the Pati-Salam model \cite{pati1974lepton,pati1975erratum} or its adapted versions \cite{mohapatra1980local,davidson1979b} based on the group $SU(4)_c \times SU(2)_L \times SU(2)_R$. These models are characterized by treating quarks and leptons on the equal footing. For instance, before $SU(4)_c$ breaking, the right-handed quarks and leptons of the first generation, which transform as a singlet under $SU(2)_L$, can be arranged into the same doublet under $SU(2)_R$ (see e.g. Refs. \cite{mohapatra1980local,babu2009neutrino,patra2014post}):
\begin{equation}
\begin{split}
\psi_R 
&=
\left(
\begin{array}{cccc}
u_1&u_2&u_3&\chi\\
d_1&d_2&d_3&e
\end{array}
  \right)_R.
\end{split}
\end{equation}
Here, the right-handed spinor is defined as $\psi_{R}\equiv P_{R} \psi$, with $P_{R}\equiv(1+ \gamma^5)/2$ being the right-handed chiral projection operator. $\chi_R$ denotes the right-handed sterile neutrino. After symmetry breaking, the right-handed fermions transform under the LRSM group in the following way \cite{mohapatra1982hydrogen,mohapatra1983spontaneous},
\begin{equation}
\begin{split}
q_R\Bigl(3, 1, 2,\frac{1}{3}\Bigr) &
 =
 \left(
  \begin{array}{c}
   u\\
    d
   \end{array}
   \right)_R, \quad
 l_R\Bigl(1, 1, 2,-1\Bigr)=
 \left(
   \begin{array}{c}
    \chi\\
    e
   \end{array}
   \right)_R.
 \end{split}
\end{equation}
Under the same symmetry group, the relevant new scalar bosons can be given by \cite{mohapatra1982hydrogen,mohapatra1983spontaneous,bolton2019alternative,nieves1984analysis,chen2011type,de2019implementing}
\begin{equation}
\begin{split}
\Delta^{(R)}_q\Bigl(\bar{6},1, 3,-\frac{2}{3}\Bigr)
&=
\left(
  \begin{array}{cc}
   \frac{\Delta_{ud}}{\sqrt{2}}         &  \Delta_{dd}\\
   \Delta_{uu} & -\frac{\Delta_{ud}}{\sqrt{2}}
  \end{array}
  \right)_R,
\end{split}
\end{equation}
\begin{equation}
\begin{split}
\Delta^{(R)}_l \Bigl(1,1,3,2\Bigr)
&=
\left(
  \begin{array}{cc}
   \frac{\Delta_{\chi e}}{\sqrt{2}}         &  \Delta_{ee}\\
   \Delta_{\chi \chi} & -\frac{\Delta_{\chi e}}{\sqrt{2}}
  \end{array}
  \right)_R.
\end{split}
\end{equation}
The new scalar bosons may lead to the instability of proton and nuclei (see e.g. Ref. \cite{dev2022searches}). As argued in Ref. \cite{mohapatra1983spontaneous}, additional discrete symmetry can be imposed on the corresponding Higgs potential so that the compatibility with the current experimental bounds on the proton lifetime $\tau_p \gtrsim 10^{31}$-$10^{33}$ yr \cite{tanabashi2018review} can be guaranteed.

Following Refs. \cite{mohapatra1980local,barbieri1981spontaneous,mohapatra1982hydrogen,mohapatra1983spontaneous,babu2009neutrino,babu2012coupling}, the relevant operators that are responsible for the $nn \rightarrow \bar{\chi}\bar{\chi}$ process depicted in Fig. \ref{dineutron} can be given by 
\begin{equation}
\begin{split}
O_{s}  \equiv & g_{\alpha \beta} q^{\alpha T}_{R} C^{-1} i \sigma_2 \Delta_q q^{\beta}_{R} + f_{\alpha \beta} l^{\alpha T}_{R} C^{-1} i \sigma_2 \Delta_l l^{\beta}_{R}\\
+ & f_{\Delta} \epsilon_{ikm} \epsilon_{jln} \Delta_{dd}^{ij} \Delta_{dd}^{kl} \Delta_{uu}^{mn} \Delta_{\chi \chi }\\
+  & \text{H.c.}
\end{split}
\label{fgdefinition}
\end{equation}
where $C$ denotes the charge conjugation operator. $g_{\alpha \beta}$, $f_{\alpha \beta}$ and $f_{\Delta}$ are dimensionless coupling constants. The $SU(3)_c$ indices are denoted by $i$, $j$, $k$, $l$, $m$, and $n$. The $SU(2)_R$ indices are denoted by $\alpha$, and $\beta$.

At the nucleon level, the $nn \rightarrow \bar{\chi}\bar{\chi}$ process can be effectively described by
\begin{equation}
-\mathscr{L}^{\text{eff}} \equiv Gs |\psi_q (0)|^4 \Bigl( \bar{n}^c \chi \Bigr) \Bigl( \bar{\chi}^c n \Bigr),\\
\label{lagrangian}
\end{equation}
or
\begin{equation}
-\mathscr{L}^{\text{eff}} \equiv Gs |\psi_q (0)|^4 \Bigl( \bar{n}^c n \Bigr) \Bigl( \bar{\chi}^c \chi \Bigr).\\
\label{lagrangian2}
\end{equation}
Here, the quark wave function at the origin takes the value: $|\psi_q (0)|^2 = 0.0144^{+3}_{-3} (\text{stat.})^{+21}_{-21}(\text{syst.})$ GeV$^3$ according to the lattice QCD calculations \cite{aoki2017improved}. The superscript $c$ represents charge conjugation. The scalar interaction couples states with opposite chirality and we have omitted the symbol of chirality for notational convenience. Following relevant studies (see e.g., Refs. 
\cite{mohapatra1982hydrogen,costa1982higgs,arnellos1982hydrogen,mohapatra1983higgs,deo1984hydrogen,mohapatra1983spontaneous,nieves1984analysis,alberico1985double}), the effective coupling constant $G_s$ can be given by 
\begin{equation}
G_s \simeq \frac{g_{uu} g_{dd}^2 f_{\chi \chi} f_{\Delta}}{M_{\Delta_{uu}}^2 M_{\Delta_{dd}}^4 M_{\Delta_{\chi \chi}}^2}.
\label{Gs}
\end{equation}
In our analysis, we assume that all the members of the new scalar bosons have similar masses \cite{costa1982higgs}, i.e. $M_{\Delta}\equiv M_{\Delta_{uu}}\simeq M_{\Delta_{dd}} \simeq M_{\Delta_{\chi \chi}} $ and all the coupling constants are also assumed to have similar values, i.e. $\lambda \equiv  g_{uu} \simeq g_{dd} \simeq f_{\Delta} \simeq f_{\chi \chi}$. These assumptions tend to be reasonable in the $SU(4)_c$ symmetry limit \cite{babu2009neutrino,babu2013post}. Even though there could be a hierarchy in the coupling constants and in the masses of the new scalar bosons, the above two relations can always be satisfied by adjusting the strengths of the coupling constants and the masses of the new scalar bosons so that a compatibility between these relations and the present limits on the stability of nuclei can be guaranteed as argued in Ref. \cite{mohapatra1983spontaneous,nieves1984analysis}. To some extent, the mass of the new scalar bosons $M_{\Delta}$ can be interpreted as the energy scale of new physics.

The new scalar bosons may also lead to the flavor-changing neutral current (FCNC) effects \cite{mohapatra2008diquark,babu2009neutrino,saha2010constraining,dorvsner2011limits,barr2012observable,babu2013post,arnold2013phenomenology,babu2013expectations,fortes2013flavor,patra2014post,sahoo2015scalar,addazi2015exotic,dev2015tev,kim2019correlation,fridell2021probing}. The phenomenology of the FCNC effects has been intensively studied. Since the FCNC processes are highly suppressed in the SM but relatively less suppressed in some new physics models, the measurements of such processes can provide a powerful tool to test the SM and to put severe constraints on the parameter space of new physics models \cite{mohapatra2008diquark,babu2009neutrino,saha2010constraining,dorvsner2011limits,barr2012observable,babu2013post,arnold2013phenomenology,babu2013expectations,fortes2013flavor,patra2014post,sahoo2015scalar,addazi2015exotic,dev2015tev,kim2019correlation,fridell2021probing}. As an important feature, the derived bounds from the FCNC processes are not usually imposed on a single coupling constant but instead they are imposed on the product of the coupling constants with different generations or flavors. Furthermore, the derived bounds also depend on the masses of the new scalar bosons. In addition, such bounds in the literature vary remarkably and it is difficult to compare them in view of the different choices of theoretical models and experimental data. If the masses of the new scalar bosons (i.e. the new physics energy scales) are within the range from several TeV to several $10$ TeV, which are accessible to a direct detection at the LHC or future high-energy experiments, the derived upper bounds on the product of the coupling constants can be roughly restricted in the range from the order of $10^{-4}$ to the order of 1. A relevant study shows that the coupling constants are more preferred to take the values in the range from the order of $10^{-4}$ to the order of 1 \cite{hao2020connection}. For purposes of illustration, we choose some typical values for the coupling constants in the range from $10^{-3}$ to $10^{-2}$ in this work. Such choices are roughly consistent with the FCNC constraints.

An approximate formula for the $nn \rightarrow \bar{\chi}\bar{\chi}$ transition rate can be found in Ref. \cite{goity1995bounds}. Under quasi-free assumptions, the transition rate can be further simplified as \cite{he2021eft,he2021b} 
\begin{widetext}
\begin{equation}
\begin{split}
\Gamma(nn \rightarrow \bar{\chi}\bar{\chi}) &\simeq \frac{\rho_{n}}{32 \pi S m_{n}^2} K(1,\xi,\xi)^{\frac{1}{2}} \overline{|\mathscr{M} (nn \rightarrow \bar{\chi}\bar{\chi})|}^2, \\
&=\frac{ K(1,\xi,\xi)^{\frac{1}{2}} }{32 \pi S}N_{f}\rho_{n}|\psi_q(0)|^8\Big(\frac{g_{uu} g_{dd}^2 f_{\chi \chi} f_{\Delta}}{M_{\Delta_{uu}}^2 M_{\Delta_{dd}}^4 M_{\Delta_{\chi \chi}}^2} \Big)^2  (m_n-m_{\chi})^2.
\end{split}
\label{eqrate}
\end{equation}
\end{widetext}
Here, $\rho_{n}$ is the neutron number density. The Kallen triangle function is defined as $K(x,y,z) \equiv x^2 + y^2 + z^2 -2xy-2yz-2zx$. The parameter $\xi$ is defined as $\xi \equiv m_{\chi}^2/(4 m_{n}^2)$ \cite{he2021eft}, where $m_n$ and $m_{\chi}$ are the mass of the neutron and sterile anti-neutrino, respectively. Since the sterile neutrinos (anti-neutrinos) may have a very light mass \cite{aghanim2020planck,boser2020status}, we assume the limit $m_{\chi} \ll m_{n}$ is satisfied. $S$ is a symmetry factor and takes the value $S=2$ \cite{he2021eft}. $N_f$ is a numerical factor from the squared amplitude and takes the value $N_f = 4$. In the second step of Eq. (\ref{eqrate}), the squared amplitude is evaluated by averaging over all initial spin configurations and summing over all final spin configurations:
\begin{equation}
\begin{split}
&\overline{|\mathscr{M} (nn \rightarrow \bar{\chi}\bar{\chi})|}^2 \\
=&\frac{1}{4} \left[ G_s |\psi_q (0)|^4 \right]^2  \text{Tr}\left[  (\cancel{p_1}+m_{n})  (\cancel{p_3}-m_{\chi})  \right] \\
&\times \text{Tr} \left[  (\cancel{p_2}+m_n)  (\cancel{p_4}-m_{\chi})  \right] \\
\simeq & N_f |\psi_q(0)|^8 \Big(\frac{g_{uu} g_{dd}^2 f_{\chi \chi} f_{\Delta}}{M_{\Delta_{uu}}^2 M_{\Delta_{dd}}^4 M_{\Delta_{\chi \chi}}^2} \Big)^2 m_n^2 (m_n-m_{\chi})^2.
\end{split}
\label{amplitude}
\end{equation}
Since our discussions are only valid up to the order of the magnitude, the terms with permutations of color and flavor indices and the corresponding numerical factors are omitted for simplicity of notation. Hence, the squared amplitude presented here is different from the one given in Ref. \cite{he2021eft} by a numerical factor. This omission always makes sense because we could absorb the possible numerical factors associated with permutation terms into the coupling constants by redefining $G_s$ without causing any inconsistencies with the present experimental limits. Furthermore, since the transition rate shows a very high power suppression by the mass of the new scalar bosons, the derived bounds on the mass of the new scalar bosons are insensitive to the omission of the numerical factors. Therefore, for the purpose of this study, we can safely omit such numerical factors. Note the transition rate formula in Eq. (\ref{eqrate}) was initially derived for the $^{16}$O nucleus \cite{he2021eft,he2021b}. If the magnitude of the Fermi-motion and nuclear binding effects in neutron stars is not too far from that in atomic nuclei, or if the rate for the dineutron decay only weakly depends on the energy of neutrons, Eq. (\ref{eqrate}) can also be applied to the neutron-star matter. At present, there is a lack of direct experimental information on the neutron-star interior and neutron-star matter. We will accept these assumptions unless they break down by future experimental data.

\section{Neutron star and equation of state}

In this section, we first review the structure of neutron stars and equation of state (EOS) for the neutron-star matter briefly. We emphasize our considerations in the choice of the EOS. Assuming that the neutron star is a static spherically symmetric object, the metric takes the form \cite{hartle1967slowly,tolman1987relativity,kogut2018special}
\begin{equation}
\begin{split}
ds^2 &= g_{\mu \nu} dx^{\mu} dx^{\nu} \\
     &= e^{2 \Phi(r)}dt^2 - \Big(1-  \frac{2GM}{r} \Big)^{-1}dr^2  -r^2 d\Omega^2,
\end{split}
\end{equation}
with the metric on the 2-sphere defined by
\begin{equation}
d\Omega^2 \equiv d\theta^2 + \sin^2 \theta d\phi^2.
\end{equation}
Here, $G$ is the gravitational constant. $\Phi(r)$ is a parameter associated with the $g_{00}$ component of the metric tensor and plays the role analogous to the Newtonian gravitational potential (i.e. effective gravitational potential).

In hydro-static equilibrium, the structure of the neutron star can be described by the Tolman-Oppenheimer-Volkoff (TOV) equations ($c \equiv 1$) \cite{oppenheimer1939massive,tolman1939static}:
\begin{eqnarray}\label{TOV}
\frac{dP(r)}{dr} &=& - \frac{[\epsilon(r) + P(r)][M(r) + 4\pi r^3 P(r)]}{r[r-2 G M(r)]}, \nonumber \\
\frac{dM(r)}{dr} &=& 4\pi r^2 \epsilon(r), \nonumber \\
\frac{d\Phi(r)}{dr} &=& \frac{r}{r-2GM(r)} \Big[\frac{GM(r)}{r^2} + 4 \pi GrP(r)\Big],
\end{eqnarray}   
where $M(r)$ is the mass within the radial distance $r$. $P(r)$ is the pressure and $\epsilon(r)$ is the energy density. Since the thermodynamic property and the chemical composition of the matter in neutron stars may vary greatly from the center to the surface, the internal structure of neutron stars can be divided into several internal layers or regions according to the current theories \cite{potekhin2015neutron,hansel2007neutron}. The energy density $\epsilon(r)$ in each region can be generally related to the mass density $\rho(r)$ by \cite{smith2012tolman}
\begin{equation}
\epsilon(r) \equiv (1 + c_j) \rho(r) + \frac{K_j}{\gamma_j -1} \rho(r)^{\gamma_j},
\label{epsiloneq}
\end{equation}
where $K_j$ and $\gamma_j$ are the normalization factor and the adiabatic index for the $j$-th region. The parameter $c_j$ can be determined by requiring that the energy density needs to be smoothly joined at the dividing density $\rho_{j}$ \cite{smith2012tolman}
\begin{align}
&\begin{aligned}
c_0 &=& 0,
\end{aligned}\\
&\begin{aligned}
c_j &=& c_{j-1} + \frac{K_{j-1}}{\gamma_{j-1} -1} \rho_{j}^{\gamma_{j-1} -1} - \frac{K_{j}}{\gamma_{j} -1} \rho_{j}^{\gamma_{j} -1}.
\end{aligned}
\end{align}
Based on a phenomenological analysis, the EOSs of neutron stars can be parameterized by a piecewise-polytropic model with three adiabatic indices ($\gamma_1$, $\gamma_2$ and $\gamma_3$) and one pressure ($P_1$) at the first dividing density \cite{read2009measuring}. The piecewise-polytropic model is useful in the analysis of astrophysical data \cite{oertel2017equations}. The effectiveness of the piecewise-polytropic parameterization has been explored in describing inspiralling binary neutron-star systems \cite{lackey2015reconstructing}. In this work, we use the parameterization scheme of the EOSs presented in Ref. \cite{read2009measuring}.

Due to the dineutron decay, the density of neutron star would decrease gradually. However, since the transition rate for the $nn \rightarrow \bar{\chi}\bar{\chi}$ process is so slow that neutron stars have sufficient time to adjust their matter distribution and maintain the hydro-static equilibrium (see e.g. Ref. \cite{berryman2022neutron}). In this case, the TOV equations can still hold in the presence of the dineutron decay. In this work, we employ the Fourth-Order Runge-Kutta (RK4) \cite{press2007numerical} approach to solve TOV equations under the boundary conditions: $\rho(0) \equiv \rho_c$, $P(R)\equiv 0$, where $\rho_c$ is the mass density at the center and $R$ is the radius of the neutron star.

At present, the EOSs of the neutron-star matter depend highly on theoretical assumptions about high-density matter and are not well-constrained, largely due to the lack of direct experimental information on the  interiors of the neutron star \cite{oertel2017equations}. Consequently, the numerical results would inevitably depend on the choice of EOSs. Nevertheless, the results obtained with various EOSs are in general consistent with each other up to one order of the magnitude. In this manner, the qualitative trends can still be identified \cite{berryman2022neutron} and useful constraints on the observable consequences can still be extracted.

The first step towards a reasonable choice of EOSs is to examine the maximum neutron-star mass determined by various EOSs. According to the astrophysical observations to date, the most massive pulsar is J0740+6620 with the mass of $2.08^{+0.07}_{-0.07}$ $M_{\odot}$ \cite{fonseca2021refined}. A recent astrophysical observation shows that the binary merger GW190814 contains an unknown compact object with a mass in the range 2.5–2.67 $M_{\odot}$ \cite{abbott2020gw190814}. If such a compact object is confirmed as a neutron star, a large class of EOSs for which the predicted maximum mass of neutron stars is smaller than 2.5 $M_{\odot}$ can be excluded. In practice, we solve the TOV equations numerically based on various EOSs and compare the yielded maximum masses of neutron stars. Massive neutron stars with the masses heavier than 2 $M_{\odot}$ can be predicted by several EOSs, such as SLy \cite{douchin2001unified}, WFF1 \cite{wiringa1988equation}, APR3 \cite{akmal1998equation}, ENG \cite{engvik1994asymmetric,engvik1996asymmetric}, ALF2 \cite{alford2005hybrid}, H4 \cite{glendenning1991reconciliation}, MPA1 \cite{muther1987nuclear}, MS1b \cite{mueller1996relativistic}, and etc. A more complete list of EOSs that lead to the masses of the neutron star greater than 2 $M_{\odot}$ can be found in Ref. \cite{biswas2022bayesian}. Some of these EOSs have been employed to model neutron stars with baryon-number violation \cite{berezhiani2021neutron,berryman2022neutron} and gravitational-wave emission \cite{lackey2015reconstructing,pacilio2022ranking}. Furthermore, a selection criteria for the EOSs can be found, for example, in Ref. \cite{carney2018comparing}. A Bayesian model selection based on multi-messenger observations shows that the MPA1 or APR3 EOSs can be more favorable in predicting the properties of  neutron stars, such as the radius and the dimensionless tidal deformability of neutron stars \cite{biswas2022bayesian}. In view of this, we choose the MPA1 EOS in our calculations. The MPA1 EOS is developed based on the relativistic Dirac–Brueckner–Hartree–Fock calculations and incorporates the contributions from the interactions mediated by $\pi$- and $\rho$-mesons \cite{muther1987nuclear}.

\begin{figure*}[t] 
\centering
\includegraphics[scale=0.99,width=0.99\linewidth]{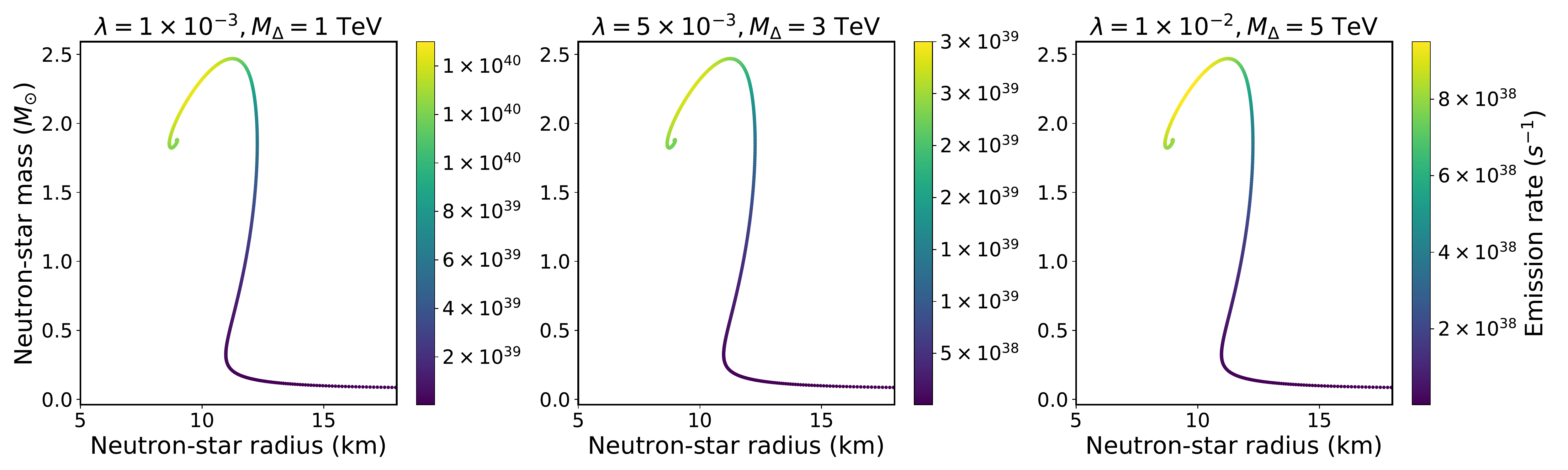}
\caption{The rate of sterile anti-neutrino emission in the mass-radius diagram of the neutron star in three different scenarios: (a) $\lambda \equiv 1 \times 10^{-3}$, $M_{\Delta} \equiv 1$ TeV; (b) $\lambda \equiv 5 \times 10^{-3}$, $M_{\Delta} \equiv 3$ TeV; (a) $\lambda \equiv 1 \times 10^{-2}$, $M_{\Delta} \equiv 5$ TeV. (Color online)
}
\label{fig2}
\end{figure*}

\begin{figure*}[t] 
\centering
\includegraphics[scale=0.99,width=0.99\linewidth]{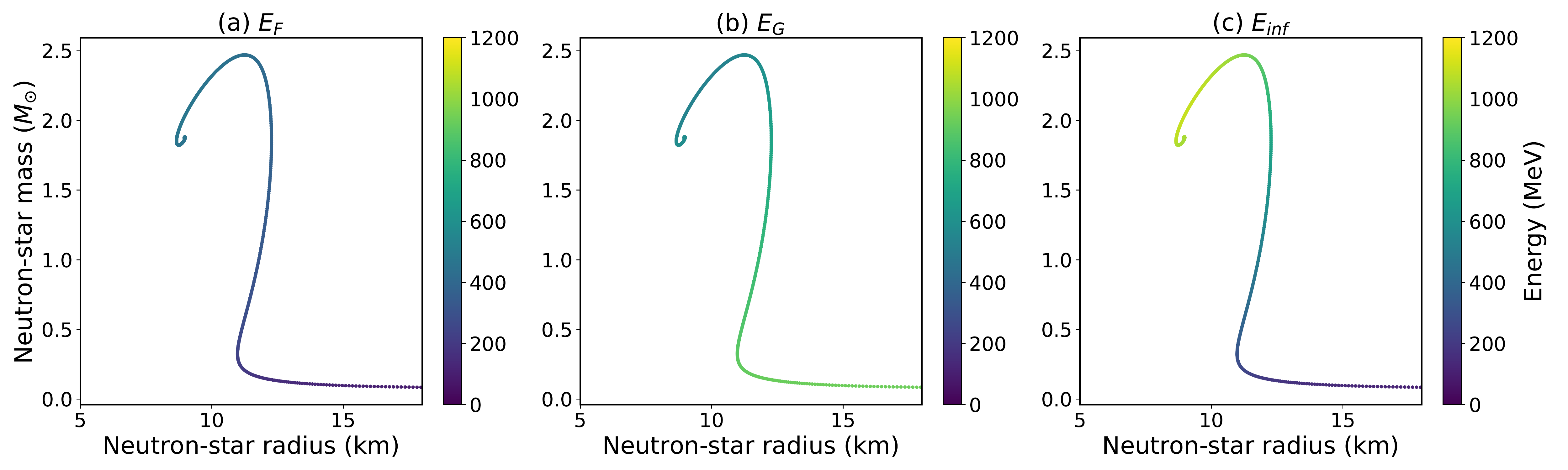}
\caption{The energy of the emitted sterile anti-neutrinos as measured from infinite and shown in the mass-radius diagram of neutron star. (a) The average Fermi energy per particle; (b) The total energy excluding the average Fermi energy per particle from infinity; (c) The total energy per particle from infinity. (Color online)
}
\label{fig3}
\end{figure*}

In what follows, we focus on the $nn \rightarrow \bar{\chi}\bar{\chi}$ process in neutron stars. We analyze the physical consequences on the properties of neutron stars and discuss about the expected signal observability at the present or future experiments.

\section{Observable consequences \label{suba}}

\subsection{Emission of sterile anti-neutrinos}
The most direct physical consequence arising from the $nn \rightarrow \bar{\chi}\bar{\chi}$ process in neutron stars is the emission of sterile anti-neutrinos. As mentioned above, since the sterile antineutrinos ($\bar{\chi}$) barely interact with ordinary matter and could be dark matter candidates \cite{abazajian2001sterile}, they can escape from the interior of the neutron star nearly without any collisions. During this process, the neutron star loses mass gradually and a large number of the sterile anti-neutrinos are emitted into space. Unfortunately, no significant evidence for the sterile neutrinos has been found so far owing to the limitations of the present experimental techniques \cite{boser2020status}. Nevertheless, these particles are expected to lie within the reach of direct searches at future high-energy experiments or future astrophysical observations. In order to develop the experimental systems more efficiently before observations and to interpret the results more correctly after observations, calculations on the properties of the emitted particles are needed.

When the emitted particles escape from the interior of the neutron star, they will lose kinetic energy to climb out of the gravitational potential and thus a gravitational red-shift occurs \cite{fuller1996neutrino}. The red-shift formula can be derived in two different ways, such as equivalence mass approach \cite{walecka2007introduction,shapiro2008black,cheng2009relativity} and frequency shift approach \cite{walecka2007introduction,shapiro2008black,cheng2009relativity}. Although the equivalence mass approach can reproduce the results of the frequency shift approach, to some extent, it is considered to be conceptually erroneous \cite{cheng2009relativity}. In the frequency shift approach, the gravitational red-shift can be evaluated by \cite{rindler2006relativity,shapiro2008black,glendenning2010special,glendenning2012compact,ferrari2020general,schutz2022first}
\begin{equation}
\eta \equiv  \sqrt{\frac{g_{00}(r_{\text{em}})}{g_{00}(r_{\text{ob}})}} = \frac{e^{\Phi(r_{\text{em}})}}{e^{\Phi(r_{\text{ob}})}}.
\end{equation}
Here, $r_{\text{em}}$ and $r_{\text{ob}}$ are the radial coordinates of the location of the emission and detection, respectively. The effective gravitational potential $\Phi(r)$ can be defined by \cite{fattoyev2010sensitivity}
\begin{equation}
\begin{split}
\Phi(r) \equiv & \int^{R}_{r}\frac{r}{r-2GM(r)} \Big[\frac{GM(r)}{r^2} + 4 \pi GrP(r)\Big]dr\\
               & - \frac{1}{2} \ln{\Bigl[1 - \frac{2 G M(R)}{R} \Bigr]}, \quad 0 < r \leq R.
\end{split}
\end{equation}

Since the sterile anti-neutrino has a negligible mass by assumption, it can escape to infinity from the interior of the neutron star. Following Refs. \cite{fuller1996neutrino,goldman2019bounds}, the energy of the sterile anti-neutrinos as measured from infinity can be defined by 
\begin{equation}
\begin{split}
E_{\text{inf}} &\simeq E_{G} + E_{F}\\
               &= \left[\frac{e^{\Phi(r_{\infty})} - e^{\Phi(r_{\text{em}})}}{e^{\Phi(r_{\infty})}}\right] m_n + \Big(\frac{3\pi^2X_nN_a}{V} \Big)^{\frac{1}{3}}\\
               &= \left[1 - e^{\Phi(r_{\text{em}})}\right] m_n + \Big(\frac{3\pi^2X_nN_a}{V} \Big)^{\frac{1}{3}},
\end{split}
\end{equation}
where $E_{F}$ is the average Fermi energy of the neutron and $E_{G}$ is the total energy excluding the Fermi energy as measured from infinity. In the second step, we choose the reference point for zero gravitational potential energy at infinity [$e^{\Phi(r_{\infty})} \equiv 1$]. We also assume that the fraction of neutrons ($X_n$) inside the neutron star approximately has the value of $0.89$ \cite{berryman2022neutron}. The total number of nucleons $N_a$ can be estimated by \cite{glendenning2012compact}
\begin{equation}
N_a \equiv 4 \pi \int^R_0 \frac{r^2 \rho_{a}(r)}{\sqrt{1-  \frac{2GM}{r} }} dr,
\end{equation}
where $\rho_{a}(r)$ is the nucleon number density.

As will be discussed below, the emission rate of the sterile anti-neutrinos and the transition rate for the $nn \rightarrow \bar{\chi}\bar{\chi}$ process depend on the coupling constants and the masses of the new scalar bosons. Here, we present some illustrative examples by choosing a few typical values for the coupling constants and the masses of the new scalar bosons. These values are generally consistent with the limits imposed by the FCNC effects as well as with the limits imposed by the observation of the binary's orbital-period changes (see Sec. \ref{subb} for more details).

Fig. \ref{fig2} shows the rate of particle emission in the mass-radius diagram of neutron stars in various scenarios corresponding to different values of the coupling constants and the masses of the new scalar bosons. As can be seen, a huge number of sterile anti-neutrinos per second ($\sim 10^{38}$-$10^{40}$ s$^{-1}$) can be emitted from the neutron star. Furthermore, the emission rate has a maximum at a specific radius and this behavior is similar to that of the neutron-star mass. This is simply due to the fact that the transition rate for the $nn \rightarrow \bar{\chi}\bar{\chi}$ process is proportional to the number density of neutrons as indicated in Eq. (\ref{eqrate}) and thus it is determined by the total number of neutrons contained in the neutron star. Due to this reason, neutron stars with a large mass and a small radius provide a more promising opportunity to search for the emitted sterile antineutrinos.

Fig. \ref{fig3} shows the energy of the emitted sterile anti-neutrinos as measured from infinity in the mass-radius diagram. As can be seen from Fig. \ref{fig3}(a) that the estimated average Fermi energy is roughly within the range from 100 to 400 MeV. The sterile anti-neutrinos gain kinetic energy from the dineutron decay and escape from the interior of the neutron star. Due to the gravitational attraction, the emitted particles lose some kinetic energies after traveling a large distance, as shown in Fig. \ref{fig3}(b). Supposing the detector is very far from the neutron star, we could use the energies at infinity to estimate the energies as measured at the location of the detector. The results are presented in Fig. \ref{fig3}(c). As can be seen, the total energy of the emitted sterile anti-neutrino as measured from infinity depends on both the radius and the mass of the neutron star. The relationships between the energy of the emitted particles and the radius and mass of the neutron star are described by multi-valued functions. Each output value in the multi-valued mappings corresponds to a different configuration of the central density $\rho_c$. Given a certain radius and mass of the neutron star, the energy spectrum of the emitted particles has a unique pattern. Quantitatively, the dineutron decay inside a neutron star is characterized by the emitted sterile anti-neutrinos with the energy from 800 to 1200 MeV as measured from infinity. From the experimental aspects, no significant evidence for such sterile particles has been observed so far. Since the number of the emitted sterile anti-neutrinos is huge, neutron stars provide an excellent opportunity for the study of such particles. If such sterile anti-neutrinos were observed, it would be a clear signal for the grand (or partially) unified models.

\begin{figure}[t] 
\centering
\includegraphics[scale=0.99,width=0.99\linewidth]{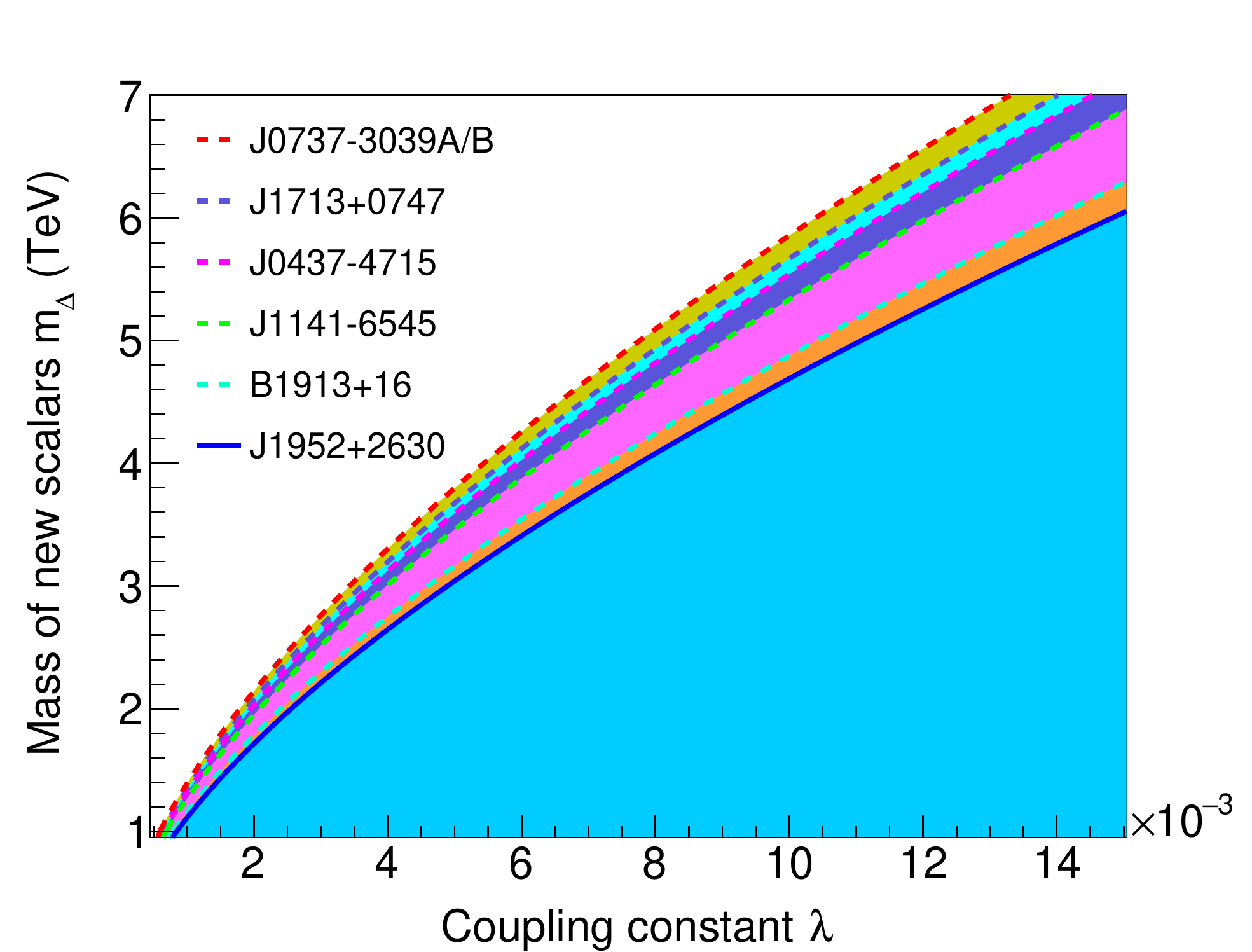}
\caption{Constraints on the mass of the new scalar bosons imposed by the observation of the binary's orbital-period changes. (Color online)
}
\label{fig4}
\end{figure}

\subsection{Orbital-period changes of binary systems \label{subb}}

\begin{table*}[t]
\caption{Constraints imposed by the observations of the binary's orbital-period changes on the mass of the new scalar bosons.}
\begin{ruledtabular}
\begin{tabular}{l|cccccccc}                                         
\diagbox{Binary sys.}{Parameter}& $M_1$ ($M_{\odot}$) & $M_2$ ($M_{\odot}$) & $|\Dot{P}/P|_{\text{BNV}}$& $|\Dot{M}/M|_{\text{BNV}}$& $^j$$M_{\Delta}$ (TeV) & $^k$$M_{\Delta}$ (TeV)\\\hline           
 J0437-4715 &  $^a$1.76 & $^a$0.254  &  $-$  & $^g$$1.6 \times 10^{-11}$&1.11 & 4.69 \\
 B1913+16   &   $^b$1.438 & $^b$1.390   &  $-$  & $^i$$6.5 \times 10^{-13}$& 1.34 & 5.67 \\
 J1952+2630 & $^c$1.35  & $^c$0.93-1.48  &  $-$  & $^g$$7 \times 10^{-12}$&1.16 & 4.88 \\
 J0737-3039A/B &  $^d$1.338185 &  $^d$1.248868 &  $^h$$7.3 \times 10^{-13}$  & $-$&1.39 & 5.86 \\
 J1713+0747 &  $^e$1.33 & $^e$0.29  &  $^h$$1.8 \times 10^{-12}$  & $-$&1.31 & 5.54 \\
 J1141-6545&  $^f$1.27 & $^f$1.02  &  $-$  & $^g$$1.6 \times 10^{-12}$&1.27 & 5.33 \\
\end{tabular}
\end{ruledtabular}
\begin{tablenotes}
\footnotesize
\centering
\item[\emph{a}]{
$^a$ Ref. \cite{verbiest2008precision}, $^b$ Ref. \cite{weisberg2016relativistic}, $^c$ Ref. \cite{lazarus2014timing}, $^d$ Ref. \cite{kramer2021strong}, $^e$ Ref. \cite{zhu2019tests}, $^f$ Ref. \cite{bhat2008gravitational}, $^g$ Ref. \cite{goldman2019bounds}, $^h$ Ref. \cite{berryman2022neutron}, $^i$ Ref. \cite{berezhiani2021neutron}; $^j$ These bounds correspond to the coupling constant $10^{-3}$; $^k$ These bounds correspond to the coupling constant $10^{-2}$; The subscript BNV denotes $\mathcal{B}$-violation.
}
\end{tablenotes}
\centering
\label{tabone}
\end{table*}

The orbital-period change of a binary system can be related to its mass change by the expression \cite{jeans1924cosmogonic}
\begin{equation}
\frac{\Dot{P}_b}{P_b} = -2 \frac{\Dot{M}}{M},
\end{equation}
where $P_b$ is the orbital period, and $\Dot{P}_b$ denotes the rate of the orbital-period change. $M$ is the total mass, and $\Dot{M}$ denotes the rate of the total mass change.

In the presence of the $nn \rightarrow \bar{\chi}\bar{\chi}$ process, the rate of the mass change for a specific neutron star $i$ ($i=1$, $2$) contained in the binary system can be approximately given by
\begin{equation}
\begin{split}
\Dot{M}_{i} &\equiv \frac{d}{dt}\int_{0}^{R(t)} 4\pi r^2 \epsilon(r, t) dr\\
& = 4\pi R(t)^2 \epsilon[R(t), t] \Dot{R}(t) + \int_{0}^{R(t)} 4\pi r^2 \Dot{\epsilon}(r, t) dr\\
& \simeq \sum_{j}\int 4\pi r^2 \Big[ (1 + c_j) \rho + \frac{K_j\gamma_j \rho^{\gamma_j}}{\gamma_j -1}\Big] \Gamma(nn \rightarrow \bar{\chi}\bar{\chi}) dr.\\        
\end{split}
\end{equation}
The summation runs over all the internal layers inside the neutron star. As mentioned above, we have also assumed that the magnitude of the Fermi-motion and nuclear binding effects in neutron stars is not too far from that in atomic nuclei and the dineutron-decay rate only weakly depends on the energy of neutrons, and thus Eq. (\ref{eqrate}) can also be applied to the neutron-star matter. In the second step, we have used the Leibniz integral rule for differentiation under the integral sign. In the last step, we have used the boundary conditions: $\epsilon[R(t), t] \simeq 0$ and $\Gamma(nn \rightarrow \bar{\chi}\bar{\chi})$ is defined by Eq. (\ref{eqrate}). Here, we have assumed that the fraction of neutrons change very slowly (i.e. $\Dot{X}_n \simeq 0$) and the following expression is satisfied:
\begin{equation}
\begin{split}
\Big\vert \frac{\Dot{\rho}}{\rho} \Big\vert \simeq \Big\vert \frac{\Dot{\rho}_n}{\rho_n} - \frac{\Dot{X}_n}{X_n} \Big\vert \simeq \Big\vert \frac{\Dot{\rho}_n}{\rho_n}\Big\vert = \Gamma(nn \rightarrow \bar{\chi}\bar{\chi}).
\end{split}   
\end{equation}

The changes in the orbital-period of the binary systems can be contributed by a number of possible sources \cite{verbiest2008precision,berezhiani2021neutron,berryman2022neutron}, such as gravitational waves, electromagnetic emission, galactic corrections, kinematic Shklovshii effects and etc. After accounting for these contributions, there are still possible discrepancies (or possible anomalous changes) that cannot be well-explained within the present theories \cite{goldman2019bounds,berezhiani2021neutron,berryman2022neutron}. Previous studies show that the possible discrepancies may mainly be attributed to the baryon-number violation (BNV) \cite{goldman2019bounds,berezhiani2021neutron,berryman2022neutron}. Motivated by these studies \cite{goldman2019bounds,berezhiani2021neutron,berryman2022neutron}, we assume that the possible discrepancies may be resulted from the $nn \rightarrow \bar{\chi}\bar{\chi}$ process. In the following discussions, we analyze the corresponding physical consequences of the $nn \rightarrow \bar{\chi}\bar{\chi}$ process on the orbital-period of the binary systems.

\begin{figure}[t] 
\centering
\includegraphics[scale=0.99,width=0.99\linewidth]{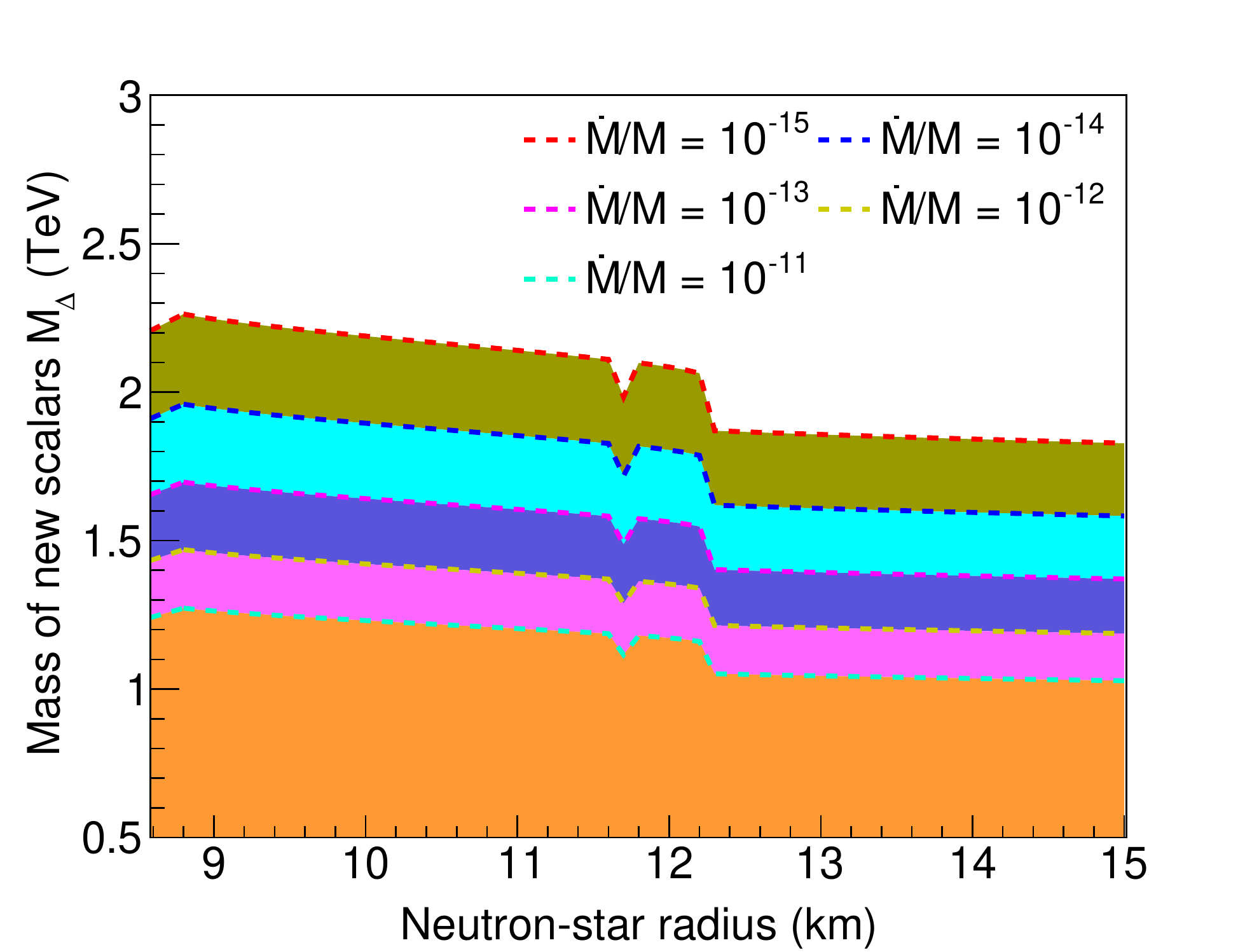}
\caption{Constraints on the mass of the new scalar bosons imposed by the mass loss of neutron star as a function of neutron-star radius in the case with the coupling constant $\lambda \equiv 10^{-3}$. (Color online)
}
\label{fig5}
\end{figure}

\begin{figure}[t] 
\centering
\includegraphics[scale=0.99,width=0.99\linewidth]{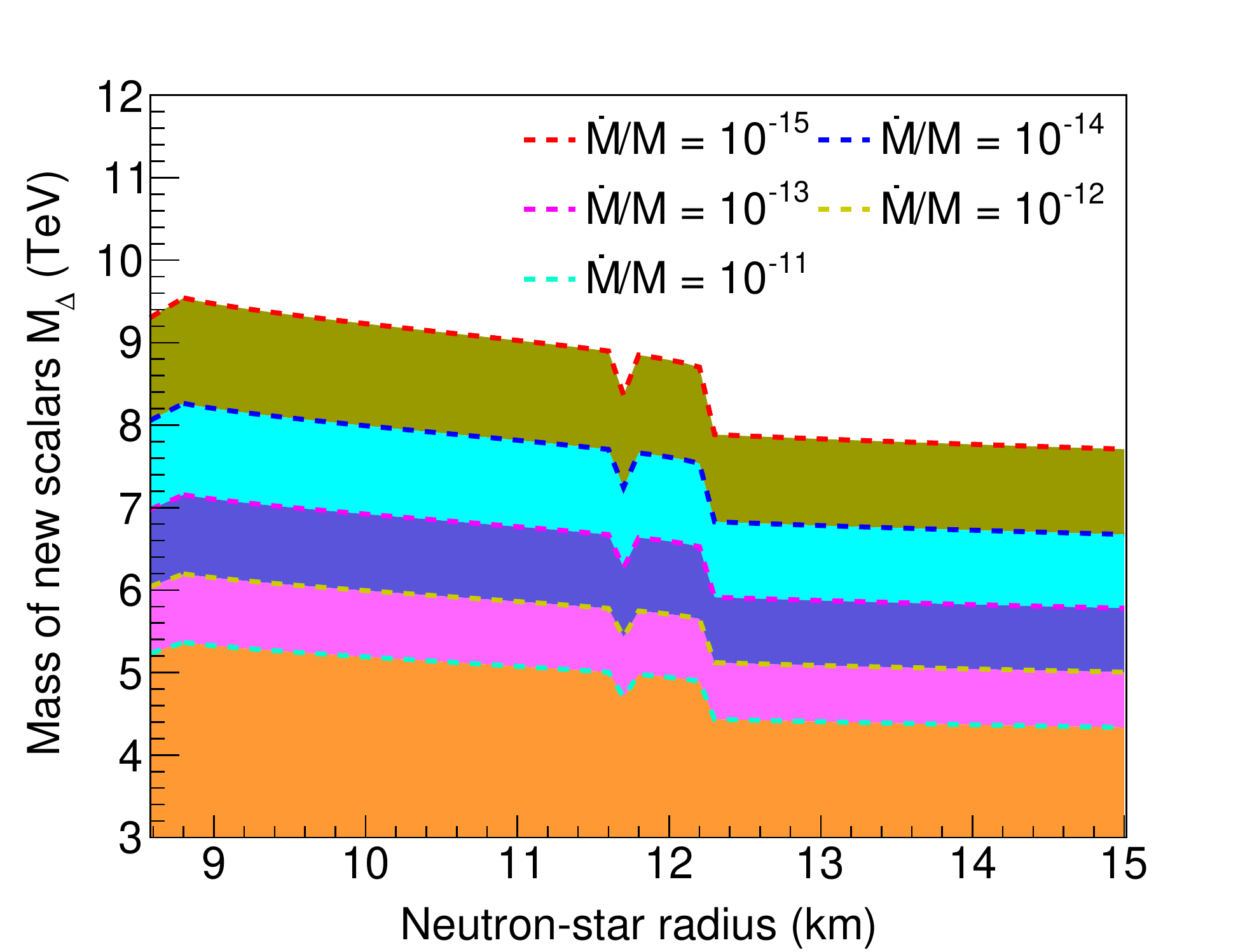}
\caption{Constraints on the mass of the new scalar bosons imposed by the mass loss of neutron star as a function of neutron-star radius in the case with the coupling constant $\lambda \equiv 10^{-2}$. (Color online)
}
\label{fig6}
\end{figure}

Tab. \ref{tabone} lists the possible discrepancies associated with the orbital-period changes or mass losses of the binary systems in previous studies \cite{goldman2019bounds,berezhiani2021neutron,berryman2022neutron} and shows the derived bounds on the mass of the new scalar bosons in the present work. The binary systems of interest in previous studies include J0437-4715 \cite{johnston1993discovery}, B1913+16 \cite{hulse1975discovery}, J1952+2630 \cite{knispel2011arecibo}, J0737-3039A/B \cite{burgay2003increased}, J1713+0747 \cite{foster1993new}, J1141-6545 \cite{kaspi2000discovery}, and etc. Among them, two binary systems, namely B1913+16 \cite{weisberg2016relativistic} and J0737-3039A/B \cite{burgay2003increased}, consist of two neutron stars. The remaining binary systems consist of a neutron star and a white dwarf \cite{verbiest2008precision,kaspi2000discovery,foster1993new,knispel2011arecibo}. The possible discrepancies associated with the relative rate of the orbital-period changes $|\Dot{P}/P|_{\text{BNV}}$ were evaluated for the binary systems, such as J0737-3039A/B ($7.3 \times 10^{-13}$ yr $^{-1}$ \cite{berryman2022neutron}), J1713+0747 ($1.8 \times 10^{-12}$ yr $^{-1}$ \cite{berryman2022neutron}), and etc. The possible discrepancies associated with the relative rate of the mass losses $|\Dot{M}/M|_{\text{BNV}}$ were evaluated for the binary systems, such as J0437-4715 ($1.6 \times 10^{-11}$ yr $^{-1}$ \cite{goldman2019bounds}), B1913+16 ($6.5 \times 10^{-13}$ yr $^{-1}$ \cite{berezhiani2021neutron}), J1952+2630 ($7 \times 10^{-12}$ yr $^{-1}$ \cite{goldman2019bounds}), J1141-6545 ($1.6 \times 10^{-12}$ yr $^{-1}$ \cite{goldman2019bounds}), and etc. Similar to the previous study \cite{foster1993new}, we assume that the $nn \rightarrow \bar{\chi}\bar{\chi}$ process only occurs inside neutron stars but does not occur inside white dwarfs. As mentioned above, the coupling constants have been shown to be roughly restricted in the range from the order of $10^{-3}$ to $10^{-2}$ \cite{hao2020connection}. For the purpose of illustration, we choose some typical values for the coupling constants in our estimation. In Tab. \ref{tabone}, the derived bounds on the mass of the new scalar bosons are presented for two different cases with the coupling constants $\lambda \equiv 10^{-3}$ and $10^{-2}$, respectively. In the case with the coupling constant of $10^{-3}$, the derived lower bounds are roughly restricted in the range from 1.1 to 1.4 TeV. In the other case with the coupling constant of $10^{-2}$, the derived lower bounds are roughly restricted in the range from 4.7 to 5.9 TeV. These bounds are higher than the existing limits reported by the direct searches at the ATLAS \cite{atlas2021search1,atlas2022search2} and CMS \cite{cms2022searches1,cms2022search2} experiments on the present LHC but could still lie within the reach of direct searches at the experiments on the upgraded LHC or future high-energy experiments.

Fig. \ref{fig4} shows the derived bounds imposed by various binary systems listed in Tab. \ref{tabone} on the mass of the new scalar bosons as a function of the coupling constants. The shaded regions have been excluded. As can be seen, the derived bounds on the mass of the new scalar bosons depend on the values of the coupling constants. A smaller coupling constant tends to give a smaller bound. The bounds arising from different binary systems are close to each other. Among them, the most stringent bound comes from the orbital-period change of the J0737-3039A/B system presented in Ref. \cite{berryman2022neutron}.

Fig. \ref{fig5} shows the constraints on the mass of the new scalar bosons imposed by the mass loss of neutron stars as a function of the neutron-star radius. Calculations are performed with the coupling constant $\lambda \equiv 10^{-3}$. Dashed lines with different colors indicate different mass losses of neutron stars, namely $|\Dot{M}/M|_{\text{BNV}} \equiv 10^{-15}$, $10^{-14}$, $10^{-13}$, $10^{-12}$, and $10^{-11}$ yr$^{-1}$. One can see that there is little change in the derived bounds with the neutron-star radius for each mass losses. In order to discern the effect of the coupling constants, the results calculated with $\lambda \equiv 10^{-2}$ are shown in Fig. \ref{fig6} for comparison. The decreasing trend in the derived bounds with respect to the neutron-star radius can still be identified. Furthermore, the steep decrease almost appears at the same neutron-star radius. This suggests that the neutron stars with radii greater than $12$ km tend to give less competitive bounds.

\begin{figure}[t] 
\centering
\includegraphics[scale=0.99,width=0.99\linewidth]{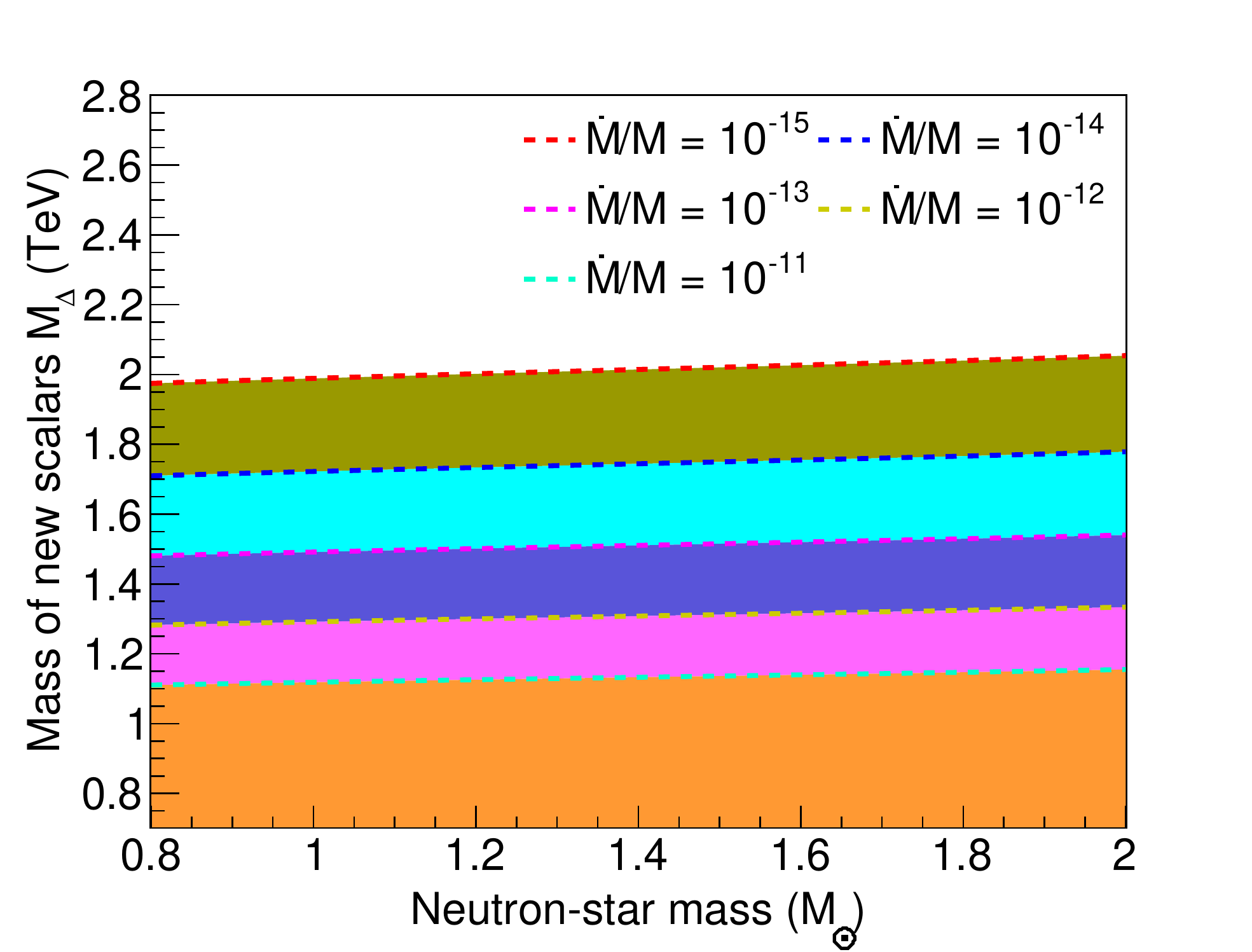}
\caption{Constraints on the mass of the new scalar bosons imposed by the mass loss of neutron star as a function of neutron-star mass in the case with the coupling constant $\lambda \equiv 10^{-3}$. (Color online)
}
\label{fig7}
\end{figure}

\begin{figure}[t] 
\centering
\includegraphics[scale=0.99,width=0.99\linewidth]{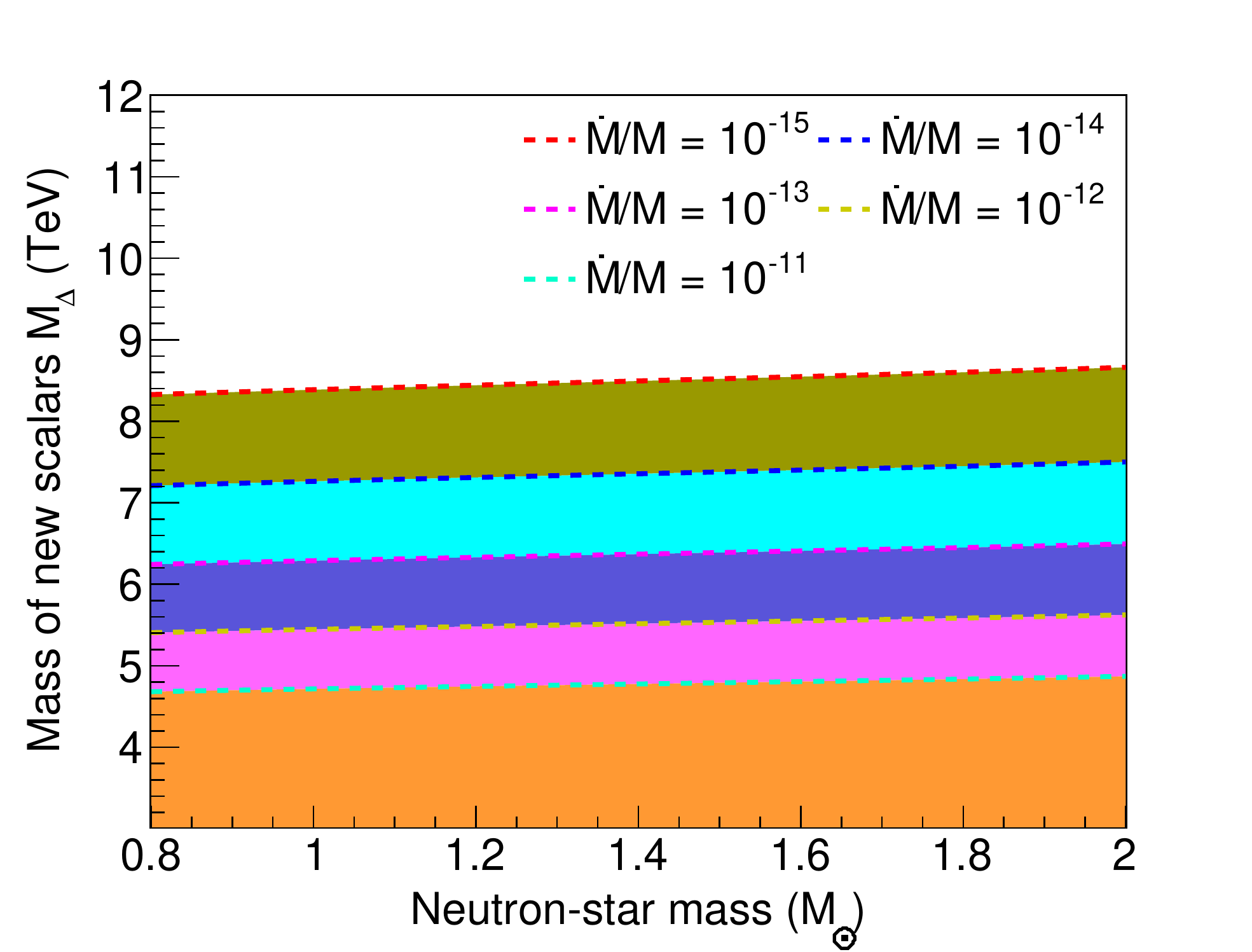}
\caption{Constraints on the mass of the new scalar bosons imposed by the mass loss of neutron star as a function of neutron-star mass in the case with the coupling constant $\lambda \equiv 10^{-2}$. (Color online)
}
\label{fig8}
\end{figure}

Figs. \ref{fig7} and \ref{fig8} show that the constraints on the mass of the new scalar bosons as a function of the neutron-star mass in two typical cases of the coupling constant $\lambda \equiv 10^{-3}$ and $10^{-2}$. For both cases, the derived bounds almost remain unchanged throughout the entire range of the allowed neutron-star masses. In contrast to the decreasing tendency with increasing the neutron-star radius, the derived bounds show an increasing trend with increasing the neutron-star mass. The neutron stars with heavier masses tend to give a more competitive bounds, but this trend is not significant.

Currently, theoretical analyses have shown that the possible discrepancies in the orbital-period changes of the binary systems roughly lie within the range from the order of $10^{-13}$ to the order of $10^{-11}$ yr$^{-1}$ (see e.g. Refs. \cite{goldman2019bounds,berezhiani2021neutron,berryman2022neutron}). On the one hand, with anticipation of upgraded experiments in astrophysics, the combined analysis between future astrophysical observations and theoretical calculations may reduce the possible discrepancies to $10^{-14}$ yr$^{-1}$ or even to $10^{-15}$ yr$^{-1}$. This would put more severe constraints on the parameter space of new physics models. On the other hand, if such discrepancies cannot be reduced and do indeed exist, they would be a clear signal of new physics.

We next analyze the compatibility between our derived bounds and the present laboratory bounds of the stability of atomic nuclei. On the one hand, similar to the statements made in Ref. \cite{berryman2022neutron}, we could assume that the $nn \rightarrow \bar{\chi}\bar{\chi}$ process can only occur, if the neutron number density exceeds a certain threshold density. Moreover, if the threshold density is lower than that in neutron stars but higher than that in atomic nuclei, the $nn \rightarrow \bar{\chi}\bar{\chi}$ process can only occur in neutron stars but cannot occur in atomic nuclei. On the other hand, we could move a step back and estimate the lifetime of atomic nuclei in the presence of the $nn \rightarrow \bar{\chi}\bar{\chi}$ process and compare it with the laboratory limits. In laboratory, the present best limit on the lifetime of the dineutron decay into invisible final states is $T_{nn \rightarrow \text{inv.}} \gtrsim  1.4 \times 10^{30}$ yr has been reported by the KamLAND collaboration based on the carbon $^{12}$C nucleus \cite{araki2006search}. More recently, another limit ($T_{nn \rightarrow \text{inv.}} \gtrsim  1.3 \times 10^{28}$ yr \cite{anderson2019search}) has been reported by the SNO+ experiment based on the $^{16}$O nucleus. Since the latter limit is weaker than the former one, we choose the limit reported by the KamLAND collaboration in the our analysis. As indicated by Eq. (\ref{eqrate}), the rate of the dineutron decay is highly suppressed by the mass of the new scalar bosons and a larger mass tends to give rise to a longer lifetime of the dineutron decay. To be conservative, we choose the mass of the new scalar bosons to be $1$ TeV, which roughly corresponds to the coupling constant $\lambda \simeq 10^{-3}$. We assume that the $^{12}$C nucleus has a spherical shape and the neutrons are uniformly distributed in the $^{12}$C nucleus. The charge radius of the $^{12}$C nucleus has the value $r_C \equiv 2.4702$ fm \cite{angeli2013table}. A rough estimate shows that the lifetime of the dineutron decay into sterile anti-neutrinos in the $^{12}$C nucleus satisfies the limit: $T_{nn \rightarrow \bar{\chi}\bar{\chi}}(^{12}\text{C}) \gtrsim 6 \times 10^{53}$ yr, which is much longer than the present experimental limits. Moreover, a larger mass would strengthen this argument. Therefore, our results are compatible with the laboratory limits and the stability of nuclei can be assured with respect to the allowed masses of the new scalar bosons.

\section{Summary}

The dineutron decay into sterile antineutrinos ($n n  \rightarrow \bar{\chi} \bar{\chi}$) violates the $\mathcal{B}$, $\mathcal{L}$, and ($\mathcal{B}+\mathcal{L}$) symmetries but conserves the ($\mathcal{B}-\mathcal{L}$) symmetry. This process is characterized by the decay of two neutrons into two back-to-back energetic sterile antineutrinos. From the theoretical aspect, this process can be mediated by the new scalar bosons and described by some new physics models with the ($\mathcal{B}-\mathcal{L}$) symmetry or the unified description of quarks and leptons. Therefore, the $nn \rightarrow \bar{\chi}\bar{\chi}$ process can serve as a promising probes for such new physics models.

Neutron stars contain a large number of neutrons and the $nn \rightarrow \bar{\chi}\bar{\chi}$ process can be significantly enhanced inside neutron stars. Due to this process, a large number of sterile anti-neutrinos can be emitted from the interior of the neutron star into space and meanwhile the neutron star loses mass and changes its properties gradually. Since the sterile antineutrinos barely interact with the ordinary matter, they can escape from the interior of the neutron star nearly without any collisions and may give rise to observable effects in astrophysical observations.

In order to estimate the impact of the $nn \rightarrow \bar{\chi}\bar{\chi}$ process on the properties of the neutron star, we have solved the TOV equations numerically based on the MPA1 EOS \cite{muther1987nuclear} using the RK4 approach \cite{press2007numerical}. The MPA1 EOS can yield a reasonable maximum mass of the neutron star and has been widely used in describing neutron stars contained in the binary systems in the literature.

In the presence of the $nn \rightarrow \bar{\chi}\bar{\chi}$ process, we have estimated the emission rate and energy spectrum of the sterile anti-neutrinos that emitted from the neutron star. The emission rate has a maximum at a specific radius and this behavior is similar to that of the neutron-star mass. Furthermore, this process is characterized by the emitted sterile anti-neutrinos with the energy from 800 to 1200 MeV. We have also pointed out that heavy neutron stars provide a more promising opportunity to observe the emitted sterile anti-neutrinos.

We have also evaluated the constraints imposed by the observations of the binary's orbital-period changes on the mass of the new scalar bosons in some typical cases of the coupling constant. In the case with the coupling constant $\lambda \equiv 10^{-3}$, the lower bounds are roughly restricted in the range from 1.1 to 1.4 TeV. In the other case with the coupling constant $\lambda \equiv 10^{-2}$, the lower bounds are roughly restricted in the range from 4.7 to 5.9 TeV. Such bounds are higher than the existing limits reported by the direct searches at the LHC but may still lie within the reach of direct searches at the upgraded LHC or future high-energy experiments.

It is expected that the combined analysis between future astrophysical observations and theoretical calculations may reduce the possible discrepancies in the binary's orbital-period changes to $10^{-14}$ yr$^{-1}$. Even a possible discrepancy as low as $10^{-15}$ yr$^{-1}$ is achievable. Such improvements might be obtained with the upgraded experiments in astrophysics and have a better chance to put more severe constraints on the parameter space of new physics models. However, if such discrepancies cannot be reduced and do indeed exist, they would be a clear signal of new physics.

\section*{Acknowledgement}
This work is supported by the National Natural Science Foundation of China (Grant No. 12022517), the Science and Technology Development Fund, Macau SAR (File No. 0048/2020/A1). The work of Yongliang Hao is supported by the National Natural Science Foundation of China (Grant No. 12104187), Macao Youth Scholars Program (No. AM2021001), Jiangsu Provincial Double-Innovation Doctor Program (Grant No. JSSCBS20210940), and the Startup Funding of Jiangsu University (No. 4111710002). Yongliang Hao thanks Dr. Yihao Yin and Dr. Leihua Liu for many useful conversations about General Relativity.

\bibliography{dineutron}
\end{document}